\documentclass[lengthcheck,twocolumn,prl,superscriptaddress,showpacs,floatfix]{revtex4-2} %tightenlines,

\usepackage{amsmath,amssymb,amsfonts,latexsym,bm}
\usepackage{graphicx,epsfig}
\usepackage{fancyvrb}
\usepackage{array}

\usepackage[svgnames]{xcolor}
\usepackage{xfrac}
\usepackage{mathrsfs}

\usepackage[colorlinks,linkcolor=MediumBlue,citecolor=MediumBlue,urlcolor = MediumBlue]{hyperref}
\usepackage{url}

\allowdisplaybreaks

\newcommand{\bu}{\hat{\bm u}}
\newcommand{\dbu}{\dot{\hat{\bm u}}}
\newcommand{\Vort}{\bm \Omega}
\newcommand{\Strain}{\bm S}
\newcommand{\ta}{t_{\rm a}}
\newcommand{\topt}{t_{\rm opt}}
\newcommand{\rmd}{{\rm d}}
\newcommand{\omegaa}{\bm \omega_{\rm a}}
\newcommand{\mycomment}[1]{}
\newcommand{\pu}{\bm p_{\hat{\bm u}}}

\begin{document}

%\title{Unraveling energetic trade-offs in microswimmer navigation: the role of body shape}
\title{Energetic cost of microswimmer navigation: the role of body shape} 

\author{Lorenzo Piro}
\altaffiliation[Current affiliation:]{~Department of Physics \& INFN, University of Rome ``Tor Vergata", Via della Ricerca Scientifica 1, 00133 Rome, Italy}
\affiliation{Max Planck Institute for Dynamics and Self-Organization (MPI-DS), 37077 G\"ottingen, Germany}
\author{Andrej Vilfan}
\affiliation{Max Planck Institute for Dynamics and Self-Organization (MPI-DS), 37077 G\"ottingen, Germany}
\affiliation{Jo\v{z}ef Stefan Institute, 1000 Ljubljana, Slovenia}
\author{Ramin Golestanian}\email{ramin.golestanian@ds.mpg.de}
\affiliation{Max Planck Institute for Dynamics and Self-Organization (MPI-DS), 37077 G\"ottingen, Germany}
\affiliation{Rudolf Peierls Centre for Theoretical Physics, University of Oxford, Oxford OX1 3PU, United Kingdom}
\author{Beno\^{\i}t Mahault}\email{benoit.mahault@ds.mpg.de}
\affiliation{Max Planck Institute for Dynamics and Self-Organization (MPI-DS), 37077 G\"ottingen, Germany}

\date{\today}

\begin{abstract}
We study the energetic efficiency of navigating microswimmers by explicitly taking into account the geometry of their body. We show that, as their shape transitions from prolate to oblate, non-steering microswimmers rotated by flow gradients naturally follow increasingly time-optimal trajectories. At the same time, they also require larger dissipation to swim. The coupling between body geometry and hydrodynamics thus leads to a generic trade-off between the energetic costs associated with propulsion and navigation, which is accompanied by the selection of a finite optimal aspect ratio.  We derive from optimal control theory the steering policy ensuring overall minimum energy dissipation, and characterize how navigation performances vary with the swimmer shape. Our results highlight the important role of the swimmer geometry in realistic navigation problems. 
\end{abstract}

\maketitle

Over the course of evolution, biological microswimmers have developed a variety of swimming mechanisms~\cite{Bray2000,Elgeti2015}.
%such as cyclic shape deformations, surface flows generated by cilia beating, or flagellar micro-rotors~\cite{}.
By means of self-propulsion, they explore their surroundings in search of food, oxygen, light, mating partners, or to escape predators \cite{Nielsen.Kirboe2021}. The energy for propulsion needs to be obtained by exploiting locally available energy sources such as light or nutrients, but their supply is often limited \cite{Mitchell2002}.
Bacterial micron-size swimmers such as {\it E. coli} use their flagella to manipulate the relative significance of translational and rotational friction \cite{Tavaddod2011} in order to control their trajectory \cite{berg}, although it has been argued that for them the metabolic cost of motion is negligible~\cite{purcell,Chattopadhyay2006PNAS,Guasto.Stocker2012}. However, larger or faster organisms such as {\it Paramecium} devote a substantial part of their energy turnover to this task~\cite{KatsuKimuraJEB2009,Taylor.Stocker2012}.
In this context, optimizing resources for navigation appears crucial for microswimmers,
while it should also be important for technological applications of artificial swimmers~\cite{tsang2020review}.

The swimming efficiency of microswimmers can be optimized by designing strategies 
that minimize the dissipated energy needed to displace the ambient fluid~\cite{Osterman2011PNAS,vilfan,Elgeti2013PNAS,guo2021,daddi-moussa-ider2021JFM,Giri2022PhysFluid}.
Such optimization problem has been the subject of recent investigation, leading in particular to the statement of several minimum dissipation theorems~\cite{babak,daddimoussaider2023minimum}.
When the swimmer moves in a nonuniform environment,
% and has access to local information about surrounding flow etc..,
a complementary approach consists in minimizing the energy dissipated along its trajectory by using, for example, the advection provided by the external flow field.
In fact, many microswimmers are equipped with receptors that allow them to measure environmental cues such as flow velocity gradient~\cite{Wheelerreview2019}, light~\cite{Jekelyphototaxis2009}, 
or concentration of certain chemicals~\cite{berg1972chemotaxis}, and use this information to navigate \cite{Bennett2015,berg}.

As the total energy spent for motion generally grows with the travel time, 
most theoretical studies on optimal navigation focus on finding time-minimizing trajectories~\cite{liebchen2019EPL,Schneider2019EPL,DaddiMoussa2021,piro2021PRR,Piro2022NJP,Piro2022frontiers,surfing,Nasiri2023EPL}, with a few exceptions~\cite{kappen2005,pinti}.
A classical example is the Zermelo problem~\cite{zermelo} in which a point-like particle moves at constant speed in a stationary flow field 
and navigates by adjusting its self-propulsion direction.
The corresponding optimal steering policy can be obtained from optimal control theory (OCT)~\cite{pontryagin,bellman} and typically depends on the local flow gradients~\cite{zermelo,liebchen2019EPL}.
Real swimmers, on the other hand, have a definite size and shape, and are thus naturally rotated by flow field gradients.
Elongated bodies like that of {\it E. coli}, for instance, undergo Jeffery rotations~\cite{jeffery} in shear flows~\cite{pedley1992hydrodynamic,RusconiNatPhys2014,JunotEPL2019}.
How the coupling between the fluid flow and the swimmer orientation affects the energy efficiency of navigation is essentially unknown.

In this Letter, we revisit the problem of optimal navigation taking into account the hydrodynamic implications of the swimmer geometry \cite{Tavaddod2011}.
We show that due to a hitherto unnoticed formal relationship between Jeffery rotations and the time-optimal Zermelo steering protocol (ZSP), 
non-navigating disk-shaped swimmers always follow minimal time trajectories. 
Considering spheroidal swimmers moving at constant speed, 
we derive from OCT the steering policy that allows them to navigate at minimal energetic cost and show that it systematically outperforms ZSP.
%which we show leads to significant improvement with respect to ZSP for elongtated bodies.
To highlight the robustness of these findings, we illustrate them in two exemplary settings of a linear shear and Gaussian-correlated random flows.

\begin{figure}[t!]
\includegraphics[width=\columnwidth]{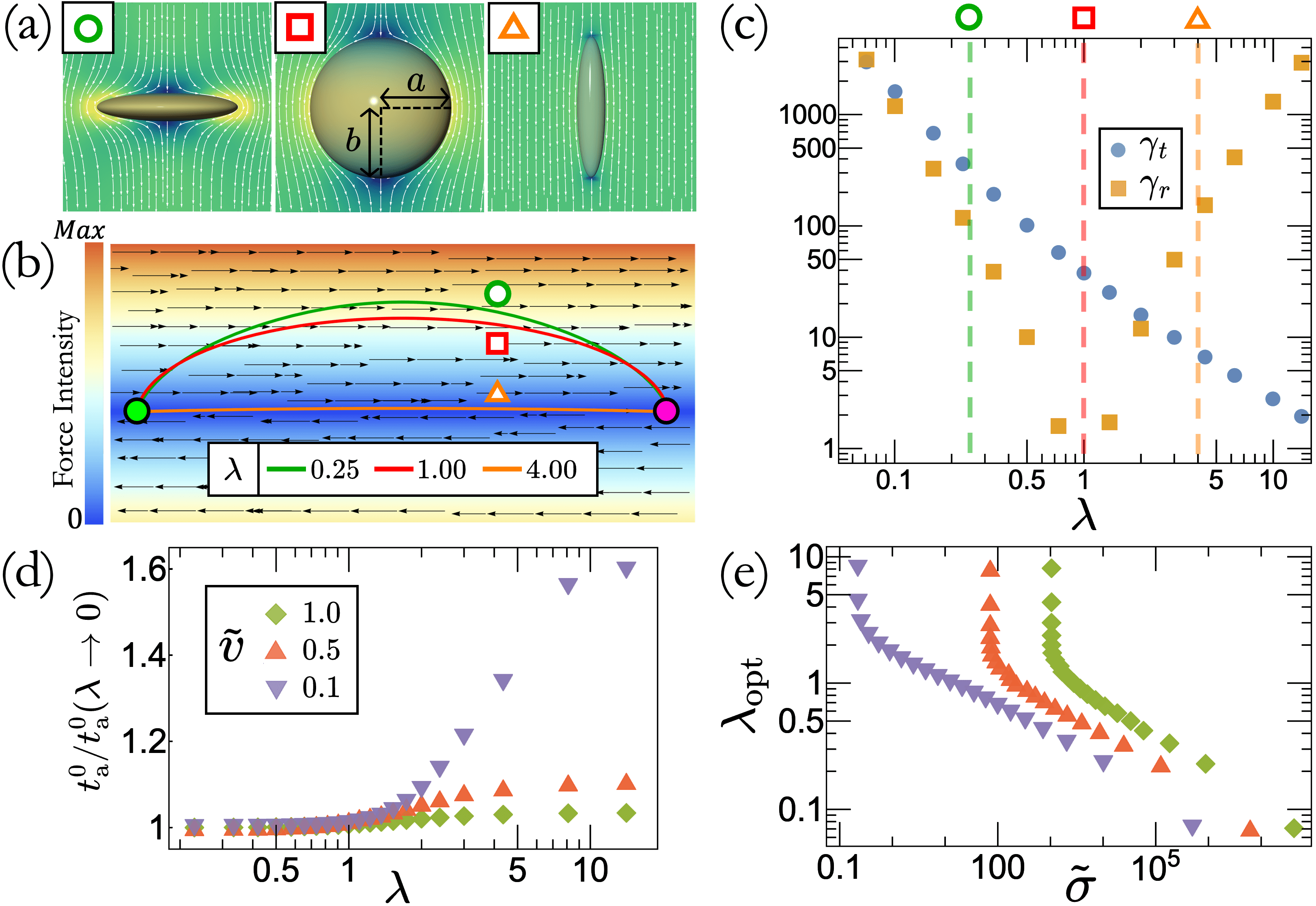}
\caption{Smart swimming by `dumb' swimmers.
(a) Flow streamlines around a (from left to right) disk-like, spherical and needle-like swimmers. 
(b) Trajectories of the non-steering swimmers 
of (a) in a linear shear flow, all starting from $\bm r_0$ (green dot) and ending in $\bm r_{\rm T}$ (magenta dot).
(c) The translational and rotational drag coefficients obtained from the minimum dissipation theorem~\cite{babak} as function of the spheroid aspect ratio $\lambda$. 
(d) Arrival time $\ta^0$ at the target as function of $\lambda$ normalized by its optimal value at $\lambda \to 0$ for three different values of $\tilde{v}$. 
(e) Optimal aspect ratio $\lambda_{\rm opt}$ as function of $\tilde{\sigma}$, legend is the same as (d).     
}
\label{fig:1}
\end{figure}

\emph{Optimal navigation with finite-size swimmers.---}
We consider a swimmer moving at constant self-propulsion speed $v_0$ in the presence of a stationary flow $\bm{f}(\bm{r})$. 
We assume this swimmer to be axisymmetric, such that its dynamics is determined by that of its position $\bm r$ and heading direction $\bu$ as
\begin{equation}
    	\dot{\bm r} = v_0\bu + {\bm f}({\bm r}), \qquad 
    	\dbu = \boldsymbol{\omega} \times \bu ,
    \label{eq:steering}
\end{equation}
where the angular velocity $\bm \omega$ comprises contributions from active torque and passive rotations. % from the flow.
Namely, $\boldsymbol{\omega} = \omegaa +\boldsymbol{\omega}_{\rm f}({\bm r},\bu)$ where $\omegaa$ is the angular velocity self-generated by the swimmer ---hereafter referred to as \emph{the control}--- while for axisymmetric bodies in low Reynolds fluids the passive contribution takes the general form
%\begin{equation} \label{eq_passive_torque}
    $\bm{\omega}_{\rm f}({\bm r},\bu) \equiv \bu \times [(\Vort + \alpha \Strain)\bu]$~\cite{jeffery,bretherton1962}.
%\end{equation}
$\Vort$ is the flow rotation and $\Strain$ the strain-rate tensor.  The coefficient $\alpha$, known as  \emph{Bretherton’s constant}~\cite{bretherton1962}, is set by the swimmer's shape.
Here, we focus on spheroidal swimmers for which $\alpha = (\lambda^2-1)/(\lambda^2+1)$~\cite{jeffery}, where the aspect ratio $\lambda \equiv b/a$ is defined such that $b$ and $a$ are the dimensions of the spheroid along and transverse to $\bu$, respectively.
As illustrated in Fig.~\ref{fig:1}(a), $\lambda < 1$ ($>1$) thus corresponds to flattened (elongated) swimmers, while spherical swimmers satisfy $\lambda=1$.

%In the classical point-to-point optimal navigation setup, also known as Zermelo's problem~\cite{zermelo}, a swimmer performance is solely evaluated against the time it takes to reach a target. 
%Remarkably, it is straightforward to show that the optimal control a swimmer should apply in that case is analogous to the passive rotation experienced by an axisymmetric particle with $\alpha=-1$ or, equivalently, by a spheroid with aspect ratio $\lambda\to0$. In other words, a disk-like particle self-propelling along its axis of symmetry will always follow time-optimal trajectories without the need to actively steer. However, the shape of a swimmer directly impacts the energy it dissipates during its motion. In fact, although a flat disk will naturally navigate faster than a thin needle, it conversely dissipates much more energy for moving.
%An interesting particular case is $\alpha = -1$, which corresponds to the Zermelo policy that minimizes travel time. For spheroids, $\alpha \approx -1$ is achievable by a flat disk propelling along its axis of symmetry.
%Considering swimmers with a well-defined shape therefore reveals a natural trade-off between travel time and energy optimization that may select a non-trivial optimal shape.

We quantify the efficiency for point-to-point navigation of a swimmer with a given shape by defining the cost
\begin{equation} 
    {\mathcal C} \equiv \int_0^{\ta}\rmd\tau \left[P_{\rm diss} + \sigma\right] \, ,
    \label{eq:cost}
\end{equation}
where $\ta$ is the total travel time. 
The first contribution to Eq.~\eqref{eq:cost} is the total energy dissipated by the swimmer during navigation, while the constant coefficient $\sigma$ is introduced to enforce travel time optimization~\footnote{Since it is dimensionally equivalent to a power, $\sigma$ can also be interpreted as the acceptable power that can be delivered by the swimmer along its trajectory.}.
In general, $P_{\rm diss}$ can be decomposed as the sum of a translational and a rotational component, $P_{\rm diss} \equiv \mu s[\gamma_t(\lambda)v_0^2 + s^2 \gamma_r(\lambda) |\bm \omega_{\rm a}|^2]$, where $s$ is a characteristic dimension of the swimmer --here defined as the radius of a sphere with equal volume-- and $\mu$ denotes the viscosity of the medium.
Here, $\gamma_t$ and $\gamma_r$ are two dimensionless coefficients that relate the dissipated power to the translational and angular swimming velocities, and take the form of effective drag coefficients. 
A lower bound on them, which is achieved with theoretically optimal propulsion, is given by the minimum dissipation theorem~\cite{babak}. This lower bound can be expressed with two drag coefficients of bodies with the same shape, one with a no-slip boundary condition and one with a perfect-slip (i.e., no tangential stress) on the surface as $\gamma_{i} = (R_{{\rm PS;}i}^{-1}-R_{{\rm NS;}i}^{-1})^{-1}$ with ${i}\in  \{ t,r\}$. For a spheroid with no-slip boundary both translational and rotational drag coefficients are known analytically~\cite{keh} while for those with perfect slip boundary we use the numerical results reported in~\cite{zwanzig}. Throughout this work, the aspect ratio $\lambda$ is varied keeping the swimmer's volume constant, such that the spheroid dimensions $a = \lambda^{-1/3}s$ and $b = \lambda^{2/3}s$.

%We have therefore generalized Zermelo's navigation problem so as to account also for the energetic efficiency related to it.
%Since we are interested in measuring how shape affects the navigation efficiency of a swimmer, let us in a first moment set the control to zero.
%Thus, optimal navigation can be performed simply via the swimmer geometry without the need of actively steering.

\emph{Smart swimming by `dumb' swimmers: the role of shape.---}
To investigate how the geometry of a microswimmer's body alone can affect its navigation performance, we first examine the case of a `dumb' swimmer that has no control over its orientation. 
For now, the active rotation $\omegaa$ is therefore set to $\bm 0$.
We consider a two-dimensional linear shear flow $\bm{f}(\bm{r})=(v_{\rm f} y/\ell)\,\bm{\hat{x}}$, where $\bm{r}=(x,y)$  and $v_{\rm f},\, \ell>0$. 
The navigation problem consists in determining the initial orientation $\bu(0)$ that allows to travel between the points $\bm{r}_0 = \bm 0$ and $\bm{r}_{\rm T} = \ell \bm{\hat{x}}$.
Rescaling space and time as $\bm r \to \ell \bm r$ and $t \to \ell t / v_{\rm f}$, the equations of motion~\eqref{eq:steering} depend only on the 
dimensionless parameter $\tilde{v}\equiv v_0/v_{\rm f}$ and the aspect ratio $\lambda$.
Due to the absence of control, the arrival time $t^0_{\rm a}$ is fully determined by $\tilde v$ and $\lambda$. 
Hence, the  cost function~\eqref{eq:cost} in units of $\mu s \ell v_{\rm f}$ reads
%\begin{equation}
    $\tilde{\mathcal C}_0 = \ta^0(\tilde{v},\lambda)\left[\gamma_t(\lambda) \tilde{v}^2 + \tilde{\sigma}\right]$,
%    \label{eq:cost_passive}
%\end{equation}
where $\tilde{\sigma} \equiv \sigma(\mu s v_{\rm f}^2)^{-1}$.

Fixing $\tilde v$, Fig.~\ref{fig:1}(b) shows that elongated swimmers ($\lambda > 1$, orange curve) typically follow nearly straight trajectories and remain within weak flow regions.
On the other hand, decreasing $\lambda$ leads to more curved trajectories, such that disk-like swimmers ($\lambda < 1$, green curve) generally benefit from an additional boost from the flow.
As shown in Fig.~\ref{fig:1}(d), this feature leads swimmers with lower aspect ratio to reach the target faster,
while the relative growth of $\ta^0$ with $\lambda$ becomes more pronounced at lower $\tilde{v}$.

%Such system shall be solved together with the boundary conditions $\bm{r}(0)=\bm{0}$ and $\bm{r}(t^0_{\rm f}) = \bm{\hat{x}}$, where the travel time $\ta^0$ is directly determined by the swimmer aspect ratio $\lambda$. 
%The bottom panel in Fig.~\ref{fig:1}(a) shows some exemplary trajectories for swimmers with the same volume ($V = \frac{4\pi}{3} s^3$, with $s = 0.1\ell$), but with different shapes. 
%A qualitative difference already stands out. Namely, needle-like swimmers ($\lambda > 1$, orange curve) tend to travel straight to the target and hence do not benefit from any boost from the underlying flow. 
%On the other hand, as $\lambda$ decreases trajectories get more curved such that disk-like swimmers (green curve) typically make greater use of the flow and are able to reach the target faster.
%This difference becomes more pronounced for smaller values of $\tilde{v}$ (Fig.~\ref{fig:1}(b)) as slower swimmers benefit more from exploiting the flow. 
% This is in agreement with the scaling obtained from the analytical expressions for the travel times in the two limits $\lambda \to 0$ and $\lambda\to\infty$, namely $\ta^0(\lambda \to 0)/\ta^0(\lambda\to\infty) \sim \sqrt{\tilde{v}}$ (details in the Appendix). Indeed, the travel times of a thin needle and a flat disk differ significantly in the limit $\tilde{v} \to 0$. For instance, already at $\tilde{v} = 0.1$ the former takes more than $50\%$ longer to reach the target than the latter (see Fig.~\ref{fig:1}(b)).
%AV: how does this fit to the sqrt(v) argument, which would give ~3x?

In fact, it is straightforward to show from OCT that Eq.~(\ref{eq:steering}) with $\omegaa=\bm 0$ and $\lambda \to 0$ ($\alpha = -1$) corresponds to the minimum travel time policy for point-like swimmers, i.e.\ ZSP~\cite{zermelo}.
In other words, thanks to passive rotations from the flow a thin disk-shaped particle self-propelling along its axis of symmetry always follows time-optimal trajectories without the need to actively steer.
Although $\ta^0$ generally increases with the aspect ratio, 
the required power to put the swimmer into motion ---here set by the coefficient $\gamma_t(\lambda)$--- is a decreasing function of $\lambda$ (Fig.~\ref{fig:1}(c)). 
These opposing trends hence imply the existence of a finite optimal aspect ratio $\lambda_{\rm opt}$ that minimizes the overall cost of navigation~\eqref{eq:cost} in complex flows.
Consistently, for the linear shear flow considered here $\lambda_{\rm opt}$ is a decreasing (respectively increasing) function of $\tilde\sigma$ (respectively $\tilde v$), as reported in Fig.~\ref{fig:1}(e).

\begin{figure*}[t!]
\includegraphics[width=.95\linewidth]{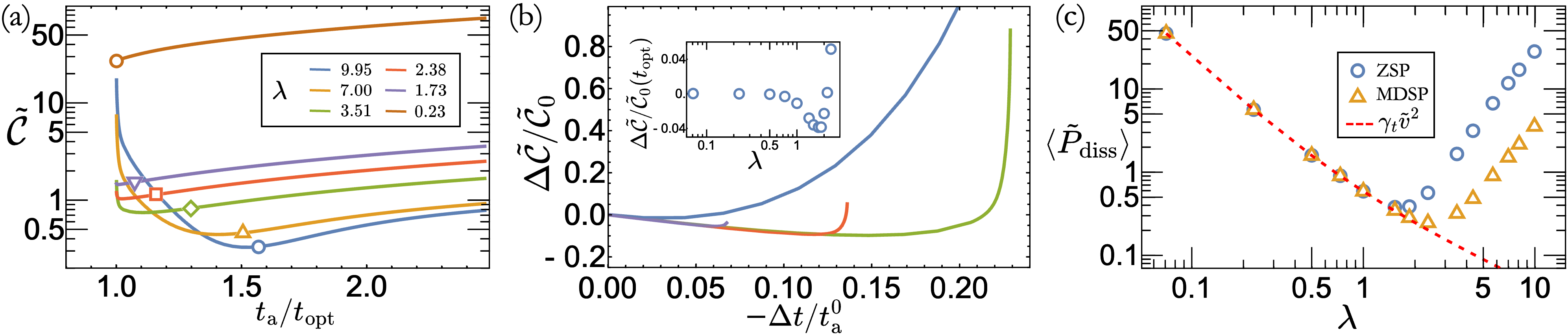}
\caption{Energy efficiency of navigating swimmers.
(a) Cost associated with optimal trajectories as function of the travel time for several values of the aspect ratio. $\ta$ is normalized by the optimal value achieved by ZSP. Symbols indicate the values $\ta^0$ and $\tilde{\cal C}^0$ obtained in absence of active steering.
(b) Relative cost variation as function of the relative travel time improvement with respect to the non-steering case (see text for definitions, same legend as in (a)). Inset: $\Delta\tilde{\cal C}$ as function of the swimmer aspect ratio for $\ta = \topt$.
(c) Comparison between the trajectory-averaged dissipated power resulting from the optimal control (orange triangles) and the compensating protocol $\omegaa^{\rm c}$ (blue circles) as function of the aspect ratio for $\ta = \topt$.
} 
\label{fig:2}
\end{figure*}

\emph{Navigation of `smart' swimmers: the cost of steering.---}
So far, we have focused on swimmers that are passively rotated by the flow and showed that their geometry alone introduces a nontrivial trade-off between energy and travel time optimization.
We now demonstrate that actively steering swimmers can achieve both energy efficient and fast navigation.
The optimal protocol for the control $\omegaa$ that minimizes the cost~\eqref{eq:cost} is determined using OCT~\cite{pontryagin,bellman}.
Defining $\bm p$ and $\pu$ as the Lagrange multipliers enforcing the equations of motion~\eqref{eq:steering},
it follows from Pontryagin’s minimization principle that the optimal value of the control 
is obtained from $\nabla_{\omegaa} {\cal H} = \bm 0$ with the effective Hamiltonian ${\cal H} \equiv P_{\rm diss} + \sigma + \bm p\cdot \dot{\bm r}+\pu\cdot\dot{\bu}$,
leading to
$\omegaa = (\pu \times \bu)/(2\mu s^3 \gamma_r(\lambda))$.
The dynamics of the momenta is in turn given by
\begin{subequations}
\label{eq_momenta}
\begin{align}
    \label{eq_momenta_p}
    \!\! \dot{\bm p} & = -\nabla_{\bm r}{\cal H} = -\nabla_{\bm r}\left[ \bm p \cdot \bm f(\bm r) + \pu\cdot(\bm \omega_{\rm f} \times \bu )\right],  \\
    \label{eq_momenta_pu}
    \!\! \dot{\bm p}_{\bu} & = -\nabla_{\bu}{\cal H} = 
    -v_0 \bm p - \nabla_{\bu}\left[ \pu \cdot (\bm \omega_{\rm f} \times \bu )\right].
\end{align}
\end{subequations}
Given a point-to-point navigation problem with unspecified initial and final particle orientations,
the minimum dissipation steering protocol (MDSP) is then obtained by integrating Eqs.~(\ref{eq:steering},\ref{eq_momenta}) with the boundary conditions $\bm r(0) = \bm r_0$, $\bm r(\ta) = \bm r_{\rm T}$, and $\pu(0) = \pu(\ta) = \bm 0$ (details about numerical methods can be found in Appendix).

For the linear shear flow setup studied above, the optimization problem additionally depends on $s/\ell$.
Since due to the control the arrival time $\ta$ can now be varied independently of $\lambda$, 
we choose it as a parameter and set $\tilde{\sigma} = 0$ without loss of generality.
Additionally, we set $\tilde{v} = \tfrac{1}{8}$ 
and $s/\ell = 0.1$.
% the characteristic swimmer size is defined by fixing its volume to $\tilde{V} = \tfrac{4\pi}{3}(s/\ell)^3$ with $s/\ell = 0.1$ when varying its shape.
%We checked that these choices do not qualitatively influence our results.
Below, we thus characterize the navigation performance of the swimmer varying the remaining two parameters $\ta$ and $\lambda$.

%We now take a step further and see how we can boost the performance of such \emph{dumb} swimmers by making them \emph{smart}. Namely, we consider swimmers which are able to navigate by actively steering their self-propulsion direction.
%, thanks to this extra degree of freedom in the swimmer dynamics,
%This essentially amounts to studying the case with $\omegaa\neq\bm0$ in Eq.~\eqref{eq:steering} and then using optimal control (OC) theory~\cite{bellman,pontryagin,bryson,bertsekas} to determine the optimal navigation policy that minimizes the total cost~\eqref{eq:cost} (details in the appendix).
%Besides, since such \emph{smart} swimmers have some control over their motion, the travel time $\ta$ is no longer predetermined by their shape, but it can rather be varied independently of $\lambda$. Consequently, the presence of $\sigma$ in the overall navigation cost~\eqref{eq:cost} does not lead to any cost variation between different shapes at fixed $\ta$ and we therefore set it to zero without loss of generality.

%Let us now discuss the performance of \emph{smart} swimmers following such optimal policy in the same exemplary shear flow setup as the one considered in the study of \emph{dumb} swimmers. 
%We shall focus on the regime of weak self-propulsion ($\tilde{v}\ll 1$) in which the performance differences between swimmers of different shapes are at largest (as shown in Fig.~\ref{fig:1}(b)). Namely, here we set $\tilde{v} = 0.125$ and checked that our results do not qualitatively change as long as $\tilde{v}<1$.

Figure~\ref{fig:2}(a) displays the dimensionless cost $\tilde{\mathcal{C}} \equiv {\mathcal{C}}(\mu s \ell v_{\rm f})^{-1}$ associated with optimal trajectories as function of the arrival time $\ta$ for several values of $\lambda$. 
As they actively steer, navigating swimmers can now reach the target in a time lower than $\ta^0$ (indicated by the symbols in Fig.~\ref{fig:2}(a)).
Remarkably, for all shapes the accessible arrival times extend to the minimum value $\topt$ achieved for $\lambda \to 0$ in absence of control.
Although for $\ta > \ta^0$ the cost decreases monotonously with $\lambda$, the regime $\ta < \ta^0$ exhibits nontrivial crossovers with elongated swimmers becoming increasingly less energy efficient at smaller times.
Hence, the optimal shape of navigating swimmers generally depends on the prescribed trajectory time.

Focusing on the regime $\ta \le \ta^0$, we show in Fig.~\ref{fig:2}(b) the relative cost variation $\Delta\tilde{\mathcal{C}} \equiv (\tilde{\mathcal{C}}-\tilde{\mathcal{C}}_0)$ associated with a 
relative travel time improvement $\Delta t\equiv (\ta-\ta^0)$.
The initial decrease of $\Delta\tilde{\mathcal{C}}$ with $-\Delta t$ attests that,
although the swimmer has to actively steer in order to reach the target in a time $\ta \lesssim \ta^0$, it does so while spending \textit{less} energy.
This feature, which we expect to hold generally, can be understood from the expression of the cost: $\tilde{\cal C} = \int_0^{\ta}\rmd\tau[\gamma_t(\lambda)\tilde{v}^2 + (s/\ell)^2 \gamma_r(\lambda)|\omegaa|^2]$.
While it is reasonable to assume that the control amplitude $|\omegaa| \simeq (\ta - \ta^0)$ such that the contribution to $\tilde{\cal C}$ of the active steering $\simeq (\ta - \ta^0)^2$,
the translational dissipation decreases linearly with $\ta$.
Therefore, so long as the swimmer travels over distances much larger than its size $(s \ll \ell)$, the cost increase resulting from active steering remains sub-dominant for $\ta \lesssim \ta^0$.
As $\ta^0 \to \topt$ for decreasing $\lambda$, oblate swimmers can then keep saving energy when optimizing their travel time down to $\topt$. 
For the navigation setup considered here, we find that such scenario occurs for $\lambda \lesssim 2$ (inset of Fig.~\ref{fig:2}(b)), 
while $\Delta\tilde{\mathcal{C}}/\tilde{\mathcal{C}}_0$ at $\ta = \topt$ exhibits a minimum at $\lambda \simeq 1.73$. 
 
%First, the mean squared angular velocity along the trajectory $\Omega_{\rm s} \equiv \int_0^{\ta}\rmd\tau |\bm \omega_{\rm a}|^2$ scales as $\sim (\ta - \ta^0)^2$ in the limit $\ta \to \ta^0$, while the translational cost is proportional to $\ta$. 
%Secondly, the rotational dissipation coefficient $\gamma_r$ scales cubically with the size $s$ of the swimmer while $\gamma_t$ scales linearly with it. Therefore, for microscopic swimmers the energy required to steer is typically much lower than the cost of propulsion.
%This can quantitatively be observed in Fig.~\ref{fig:2}(b) where we show the relative cost difference with respect to the \emph{dumb} swimming scenario, thus defined as $\Delta\mathcal{C}_\%\equiv (\mathcal{C}-\mathcal{C}_0)/\mathcal{C}_0\cdot100$, as a function of the corresponding improvement in the travel time, i.e. $\Delta t_\% \equiv (\ta-\ta^0)/\ta^0\cdot100$.
%In order to achieve shorter arrival times, slender swimmers start to use significantly more energy to navigate, resulting in a sharp increase of the cost~\eqref{eq:cost}.
%On the other hand, more spherical swimmers with aspect ratio $0.5\lesssim\lambda\lesssim2$ are able to keep saving energy even when optimizing their travel time down to $\topt$. 
%The cost saving (or excess cost) of swimmers that reach the target in the shortest time $\topt$ is displayed in the inset of Fig.~\ref{fig:2}(b). The saving is maximal for a swimmer with the aspect ratio $\lambda\simeq1.73$. 

As MDSP is able to provide the minimum time trajectories naturally followed by disk-like swimmers, it is instructive to compare it with a naive implementation of ZSP.
Namely, we consider a control $\omegaa^{\rm c}(\lambda) \equiv \bm\omega_{\rm f}(\lambda = 0) - \bm\omega_{\rm f}(\lambda)$ 
that compensates for the shape-dependent hydrodynamic rotations and implements the steering protocol  minimizing travel time for a point-like swimmer.
As shown in Fig.~\ref{fig:2}(c), for slender swimmers following MDSP the trajectory-averaged dissipated power $\langle \tilde{P}_{\rm diss} \rangle \equiv \tilde{\cal C}/\ta$ is about an order of magnitude lower than that associated with the simpler compensating control $\omegaa^{\rm c}$.
On the other hand, for oblate swimmers which barely steer $\langle \tilde{P}_{\rm diss} \rangle \approx \gamma_t \tilde{v}^2$ in both cases.

%As a naive guess one could have expected the optimal navigation policy that allows swimmers of any shape to reach the target in the shortest time to be a trivial generalization of ZSP. 
%This would effectively amount to offsetting the flow-induced torques while adding back Zermelo's solution, which simply translates into imposing $\omegaa(\lambda) = \bm\omega_f(\lambda\to0) - \bm\omega_f(\lambda)$ in \eqref{eq:steering}.
%However, as shown in Fig.~\ref{fig:2}(c), the dissipated power $P_{\rm diss}$ of this strategy turns out to be systematically larger than the one resulting from the implementation of the policy obtained from OC theory. Remarkably, smart swimmers with any aspect ratio following the latter strategy thus succeed in reaching the target in the theoretically optimal time whilst spending less energy than required by ZSP.
%Hence, we find that the latter policy effectively outperforms ZSP. 

\emph{Navigation in a complex environment.---}
So far, the analysis of the navigation performance was carried out in a simple linear shear flow.
We now highlight the generality of the above results by considering a stationary two-dimensional flow defined from a random stream function $\psi(\bm r)$ having zero mean and Gaussian correlations~\cite{Gustavsson2016}: $\langle\psi(\bm{r})\psi(\bm{r'})\rangle = \tfrac{1}{2}\ell^2 v_{\rm f}^2 e^{ -|\bm{r}-\bm{r'}|^2/(2\ell^2) }$.
The parameters $\ell$ and $v_{\rm f}$ hence correspond to the correlation length and mean intensity of the random flow.
For the simulation results shown below, a single instance of the random flow in a periodic square domain of length $L = 25\ell$ was generated (details in Appendix). 

%Let us now highlight the generality of the results that were so far illustrated in the simple shear flow setup. To this end, we consider a two-dimensional stationary Gaussian random flow field defined as $\bm{f}(\bm{r}) \equiv \frac{1}{\sqrt{2}}\bm{\nabla}\times[\bm{\hat{e}}_z \psi(\bm{r})]$. Here $\bm{\hat{e}}_z$ denotes the unit-vector in $z$-direction and $\psi(\bm r)$ a random stream function with zero mean and correlation function~\cite{Gustavsson2016}: $\langle\psi(\bm{r})\psi(\bm{r'})\rangle = \ell^2 v_{\rm f}^2 \exp\left[ -\frac{|\bm{r}-\bm{r'}|^2}{2\ell^2} \right]$, where $\ell$ and $v_{\rm f}$ are the characteristic length and flow intensity scales, hereafter set to $\ell=1$ and $v_{\rm f}=1$. We numerically generate an instance of the random flow in a square box $L\times L$ with periodic boundary conditions (more details in the appendix). 

\begin{figure}[t!]\centering
\includegraphics[width=\columnwidth]{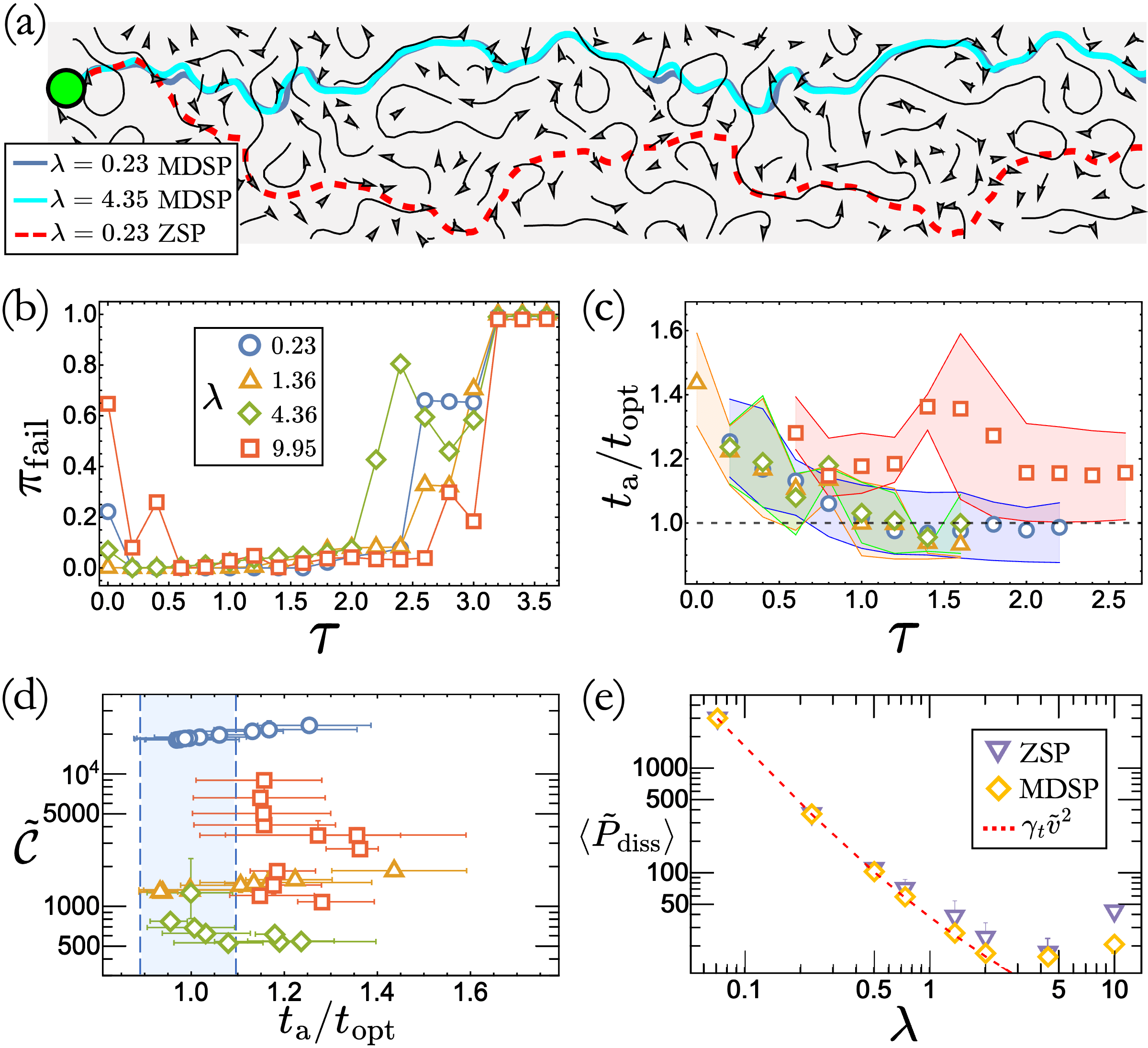}
\caption{Optimal navigation in a Gaussian random flow.
(a) Three exemplary trajectories with travel time $\ta \approx \topt$ starting from the same point (green dot) and obtained from the local approximations of ZSP (dashed red curve) and MDSP (solid curves).
The displayed domain size is $2 L \times L/2$ and the black arrows show the local direction of the flow.
(b) Probability of not reaching the finish line as function of $\tau$ for four different aspect ratios $\lambda$.
(c) Mean travel time $\ta$ as function of $\tau$.
The shaded intervals show the first and third quartiles
while the dotted line indicates the reference value $\topt$.
%dotted line indicates the mean time taken by a swimmer following the approximated version of ZSP.
(d) Mean cost of navigation as function of the arrival time. 
Error bars show the first and third quartiles, 
while the shaded area indicates the interval of confidence on $\topt$. 
Legends for (c,d) are the same as (b).
(e) Comparison between the average dissipated power of the approximated MDSP and ZSP at arrival time $\ta \approx \topt$ as function of the swimmer aspect ratio. 
%Comparison between the dissipated powers of the two strategies derived by locally approximating ZSP and the OC policy when reaching the target in a time $\ta \simeq \topt$ as function of the swimmer aspect ratio.
}
\label{fig:3}
\end{figure}

We also study a more realistic navigation problem requiring the swimmer to travel from an initial position on a vertical line at $x = 0$ to a finish line located at $x = 4L$ (see Fig.~\ref{fig:3}(a)).
In order to account for this new navigation task, we revised the cost function as
%\begin{equation}
 ${\mathcal C}_{\rm RF} \equiv -\kappa (x(\ta) - x(0)) + \int_0^{\ta} \rmd\tau\, P_{\rm diss}$,
    %\label{eq:cost2}
%\end{equation}
where the $\propto \kappa$ boundary term is intended to maximize the drift towards the finish line.
Using $\ell$ and $\ell/v_{\rm f}$ to rescale space and time, we are left with 4 parameters: $\lambda$, $\tilde{v}$, $s/\ell$, and $\tilde{\kappa} \equiv \kappa(\mu s v_{\rm f})^{-1}$.
According to OCT, the revised MDSP for a given initial position $\bm r_0$ is obtained solving the system of differential equations~(\ref{eq:steering},\ref{eq_momenta}) with boundary conditions $\bm r(0) = \bm r_0$, $\bm p(\ta) = -\kappa \hat{\bm x}$, and $\pu(0) = \pu(\ta) = \bm 0$.
However, in practice solving such boundary value problem using e.g., shooting methods, in a complex environment turns out to be computationally unfeasible due to the chaotic nature of the solutions~\cite{Piro2022frontiers,biferale}. 

Reinforcement leaning-based approaches have become increasingly popular to tackle complex navigation scenarios~\cite{colabrese,Schneider2019EPL,biferale,muinos2021ScienceR,Mousavi2022,Nasiri2022,Nasiri2023EPL,Putzke2023EPJE}.
Here, we instead designed an alternative approach that locally approximates the optimal control policy.
As detailed in the Appendix, we define a time horizon $\tau$ which we assume sufficiently small such that $\bm r$ and $\bu$ do not significantly vary over the interval $[0;\tau]$.
Within this assumption, Eqs.~\eqref{eq_momenta} reduce to a linear system of differential equations which we can solve exactly.
Given a swimmer with specified position and orientation at time $t$, we thus solve the optimization problem over the interval $[t ; t + \tau]$ by determining the values of $\bm p(t)$ and $\pu(t)$ from the solution of~\eqref{eq_momenta} and the conditions $\bm p(t + \tau) = - \tilde{\kappa} \hat{\bm x}$ and $\pu(t + \tau) = \bm 0$.
We obtain this way an approximation of the optimal control $\omegaa(t)$ that minimizes the navigation cost at fixed $\tau$.
In practice, the optimal value of $\tau$, whose existence can be rationalized, is determined empirically.
The limit $\tau \to 0$ indeed amounts to ignoring the presence of the flow field, such that it leads the swimmer to point straight to the target line.
Conversely, for large $\tau$ the approximation of constant couplings in Eqs.~\eqref{eq_momenta} becomes increasingly poor.
Similar optimization protocols have been implemented in Refs.~\cite{surfing,calascibetta2023optimal}.
The particularity of the approach we propose is that it is generalizable to any optimal control problem sharing the same Hamiltonian structure as Eqs.~(\ref{eq:steering},\ref{eq_momenta}).
In fact, applying it to the Zermelo problem we recover the protocol derived in~\cite{surfing}.

As they do not strongly influence the results, we set $\tilde{v} = 1$, $s/\ell = 0.1$ and $\tilde{\kappa}\tau\tilde{v} = 10^2$. 
%and checked that reasonable variations around this value lead to qualitatively similar results.
Figure~\ref{fig:3}(a) shows representative trajectories obtained from the local approximations of MDSP and ZSP.
In both cases they end up on an attractor after crossing typically one or two system sizes $L$ along the $x$ direction.
The number of reachable attractors depends on the policy employed.
To account for all of them, all data was collected by averaging over the initial swimmer position $\bm r_0 = (0,y_0)$ with $y_0$ uniformly distributed in $[0;L]$.

Since for some parameters swimmers might get trapped in strong flow regions,
we define $\pi_{\rm fail}$ as the probability that a swimmer does not reach the finish line in a finite simulation time.
As shown in Fig.~\ref{fig:3}(b), for MDSP $\pi_{\rm fail}$ exhibits a minimum in the range 
$0.5 \lesssim \tau \lesssim 2$ for all values of $\lambda$. 
Measuring the average arrival time of trajectories that successfully reach the goal, we find that it is generally minimal within a similar range of $\tau$ values (Fig.~\ref{fig:3}(c)). 
A similar trend is observed for ZSP, and we use the corresponding minimum arrival time $\topt \approx 0.7\times4L(\ell\tilde{v})^{-1}$ as a reference value.
Consistently with results obtained in the simple shear flow, for most aspect ratios there exists a value $\tau \approx 1$ for which $\ta$ obtained from MDSP is comparable to $\topt$ (Fig.~\ref{fig:3}(c)).
Comparing Figs.~\ref{fig:3}(d,e) with Figs.~\ref{fig:2}(a,c), the results obtained in the shear flow are qualitatively confirmed.
Namely, we observe the selection of finite optimal aspect ratio set by the arrival time.
Moreover, for $\ta \approx \topt$ MDSP systematically performs better than ZSP while the two converge as $\lambda \to 0$ where they satisfy $\langle \tilde{P}_{\rm diss}\rangle \approx \gamma_t(\lambda)\tilde{v}^2$.

We have introduced a general formalism to study the influence of a microswimmer's body geometry on its navigation performances.
Our analysis reveals that, due to the connection between ZSP and rotations of disk-shaped particles in flows,  
non-steering swimmers typically travel faster the lower their aspect ratio $\lambda$.
As the dissipated energy follows the opposite trend, 
a nontrivial value of $\lambda$ that minimizes the overall navigation cost is generally selected.
These features have interesting consequences for actively steering swimmers such as the possibility to simultaneously decrease travel time and dissipated energy by navigating,
or the existence of a 
travel-time dependent optimal aspect ratio.
Interestingly, MDSP is distinct from a simple generalization of ZSP which systematically leads to higher energy dissipation.
Given the generality of our approach, we expect these results to be applicable to a broad range of systems, such that they may be helpful for the design of smart artificial swimmers~\cite{tsang2020review}.

\acknowledgments
We thank Tarun Mascarenhas for discussions at the early stages of this work.
LP is thankful for the funding from the International Max Planck Research School (IMPRS) for the Physics of Biological and Complex Systems.
%, and for the partial support from the European Research Council (ERC) under the European Union’s Horizon 2020 research and innovation programme (Grant Agreement No.\ 882340).
AV acknowledges support from the Slovenian Research Agency (Grant No.\ P1-0099).
This work has received support from the Max Planck School Matter to Life and the MaxSynBio Consortium, which are jointly funded by the Federal Ministry of Education and Research (BMBF) of Germany, and the Max Planck Society.\\

\appendix

\setcounter{figure}{0}
\renewcommand\thefigure{A\arabic{figure}}
\setcounter{equation}{0}
\renewcommand\theequation{A\arabic{equation}}

\noindent\emph{Appendix on the numerical solutions of the optimization problems.---}
Let us consider a generic navigation problem where the swimmer state is parameterized by the vector $\bm{q}(t) = (q_1(t),\ldots,q_n(t))$ whose deterministic evolution follows $\bm{\dot{q}} = \bm{S}[\bm{q}(t),\bm{c}(t),t]$, where $\bm{c}$ denotes the control that can be used for navigation.
The navigation task consists in finding the trajectory minimizing the cost function
\begin{equation}
    {\cal C} = \phi(\bm{q}(\ta),\ta) + \int_{0}^{\ta} \mathcal{L}[\bm{q}(t),\bm{c}(t),t] \, {\rm d}t \ , 
    \label{eq:cost_OC}
\end{equation}
with boundary conditions $q_i(0) = q_{i,0}$ 
and $q_j(\ta) = q_{j,\rm{T}}$, where the indices $1 \le i \le k$ and $m \le j \le l$ such that for $k,l < n$ or $m > 1$ certain degrees of freedom can be unspecified at the two ends of the trajectory.
$\phi$ and $\mathcal{L}$ in~\eqref{eq:cost_OC} are respectively known as the endpoint and running costs.
OCT recasts this optimization problem into a boundary value problem for the dynamical system~\cite{bryson}
\begin{equation} \label{eq_APP_Hamiltonian}
    \bm{\dot{q}} = \bm{\nabla_p}\mathcal{H}, \qquad
    \bm{\dot{p}} = -\bm{\nabla_q}\mathcal{H},
\end{equation}
where the Hamiltonian $\mathcal{H}\equiv\mathcal{L}[\bm{q}(t),\bm{c}(t),t]+\bm{p}(t)\cdot\bm{S}[\bm{q}(t),\bm{c}(t),t]$ and with the boundary conditions 
\begin{align*}
        q_i(0) & = q_{i,0} & 1 & \le i \le k \; {\rm (specified)}, \\
        p_i(0) & = 0 & k & < i \le n \; {\rm (unspecified)}, \\
        q_j(\ta) & = q_{j,\rm{T}} & m & \le j \le l \; {\rm (specified)}, \\
        p_j(\ta) & = \partial_{q_j}\phi & 1 & \le j < m \wedge l < j \le n \; {\rm (unspecified)} .
\end{align*}
In addition, the navigation policy for the control is obtained by minimizing the Hamiltonian: $\bm{\nabla_c}\mathcal{H} = 0$.
%Should we say something about $\ta$?
%Finding the optimal trajectory thus requires to determine the unknown $k$ components of $\bm p(0)$ and unspecified $n-k$ components of $\bm q(0)$.

\begin{table}[!t]
    \centering
    \caption{Correspondence between the general formulation of optimal navigation and the two policies studied in the main text.}
    \begin{tabular}{p{0.34\columnwidth}>
    {\centering}p{0.31\columnwidth}
    {c}p{0.34\columnwidth}}
    \hline\hline
        Problem & ZSP & MDSP \\
        & (min. time) & (min. dissipation) \\
        \hline
        State variables $(\bm q)$ & $\bm r$ & $\bm r$, $\bu$ \\
        Conjugate variables & $\bm p \leftrightarrow \bm r$ & $\bm p \leftrightarrow \bm r$,  $\; \pu \leftrightarrow \bu$ \\
        Optimal control $(\bm c)$ & $\bu = -\bm p / |\bm p|$ & $\omegaa \propto (\pu \times \bu)$ \\
        Running cost $({\cal L})$ & $\sigma$ & $P_{\rm diss} + \sigma$\\
    \hline\hline
    \end{tabular}
    \label{tab:policies}
\end{table}

The correspondence between the general optimization problem and the two navigation protocols addressed in the text is summarized in Table~\ref{tab:policies}.
For the study of navigation in linear shear flow, we solved the boundary value problem via standard shooting methods.
Namely, given a trajectory time $\ta$ and a guess for the $n$ unknown initial conditions $(\{q_i(0)\}_{i\in[k+1;n]};\{p_i(0)\}_{i\in[1;k]})$, the coupled systems of ordinary differential equations of~\eqref{eq_APP_Hamiltonian} are integrated via the $4^{\rm th}$ order Runge-Kutta method with time step ${\rm d}t=10^{-5}$.
The initial conditions are then iterated using the routine \texttt{gsl\_multiroot\_fsolver\_hybrids} provided by the GSL library~\cite{gough2009} to determine the roots of the system 
%\begin{equation*}
$(\{q_j(\ta) - q_{j,T}\}_{j \in [m;l]};\{p_j(\ta)-\partial_{q_j}\phi\}_{j\in[1;m-1]\cup[l+1;n]})$.
%\end{equation*}
This process is then iterated until reaching convergence, which
we define as when the sum of absolute errors falls under a specified threshold (here set to $10^{-6}$).\\

\noindent 
\emph{Appendix on the approximate navigation policies.---}
Here, we give additional details about the derivation of the approximate navigation policies described in the text.
To keep the presentation simple, we restrict the problem to two dimensions for which Eqs.~(\ref{eq:steering},\ref{eq_momenta}) simplify as 
\begin{subequations}
\begin{align}
\label{eq:app_motion2D_r}
    \bm{\dot{r}} & = v_0\bu(\theta) + {\bm f}({\bm r}) \\ 
    \label{eq:app_motion2D_theta}
    \dot{\theta} & = \omega_{\rm a} + \omega_{\rm f}(\bm{r},\theta) \\
    \label{eq:app_motion2D_p}
    \dot{\bm p} & = -\nabla_{\bm r}\left[ \bm p \cdot \bm f(\bm r) + p_{\theta} \omega_{\rm f}(\bm{r},\theta)\right],  \\
    \label{eq:app_motion2D_ptheta}
    \dot{p}_{\theta} & = 
    - v_0 \bm p \cdot \bu^{\perp}(\theta) - \partial_{\theta}\left[ p_{\theta} \omega_{\rm f}(\bm{r},\theta) \right],
\end{align}
\label{eq:app_motion2D}
\end{subequations}
where $\bu^{\perp}(\theta) \equiv \rmd \bu(\theta)/\rmd\theta$ and $\omega_{\rm a} = -{p}_\theta(2\mu s^3 \gamma_r)^{-1}$.

Given a swimmer with position $\bm r$ and orientation $\theta$ at time $t$, we wish to determine the control $\omega_{\rm a}(t)$ that minimizes the cost
${\mathcal C}_{\rm RF} = -\kappa (x(\ta) - x(t)) + \int_t^{\ta} \rmd\tau\, P_{\rm diss}$.
From OCT, the optimal control is obtained solving the boundary value problem for the system~\eqref{eq:app_motion2D} with the end-point conditions  $\bm p(\ta) = -\kappa \hat{\bm x}$ and $p_{\theta}(\ta) = 0$.
Since the solutions of~\eqref{eq:app_motion2D} are generally chaotic in presence of strong or complex flows~\cite{Piro2022frontiers}, shooting methods do not necessarily converge.
Instead, we derive an approximation for $\omega_{\rm a}(t)$ that relies only on the information locally available to the swimmer.

Denoting $\textbf{P} \equiv (\bm p, p_{\theta})$, the dynamics of the conjugate variables can be written as $\dot{\textbf{P}} = - \textbf{M}\cdot\textbf{P}$ where the time-dependent coefficient matrix reads
\begin{equation*}
    \textbf{M} \equiv \begin{pmatrix}
                    {\rm F}_{11} & {\rm F}_{12} & \Omega_1 \\
                    {\rm F}_{21} & {\rm F}_{22} & \Omega_2 \\
                    -v_0\sin\theta & v_0\cos\theta & \Omega_3
                \end{pmatrix} \, ,
\end{equation*}
with
\begin{equation*}
    \textbf{F} \equiv \begin{pmatrix}
                    \partial_xf_x & \partial_xf_y \\
                    \partial_yf_x & \partial_yf_y
                \end{pmatrix} \, , \qquad
    \bm{\Omega} \equiv \begin{pmatrix}
                    \partial_x\omega_{\rm f} \\
                    \partial_y\omega_{\rm f} \\
                    \partial_\theta\omega_{\rm f} 
                \end{pmatrix} \, .
\end{equation*}

We now assume that the variations of $\bm r$ and $\theta$ are sufficiently smooth such that there exists a timescale $\tau$ over which the coefficients of the matrix $\textbf{M}$ are nearly constant.
Under this assumption, the solution for $\textbf{P}$ is readily obtained for $t' \in [t; t + \tau]$ as
\begin{equation}
    \textbf{P}(t') \simeq e^{-(t'-t)\textbf{M}}\cdot\textbf{P}(t) \, ,
    \label{eq:OC_P}
\end{equation}
where the coefficients of the matrix $\textbf{M}$ are evaluated at time $t$.
Optimizing the cost ${\mathcal C}_{\rm RF}$ over this time window then imposes that 
${\rm P}_i(t + \tau) = -\kappa\delta_{i,1}$ which, 
together with Eq.~\eqref{eq:OC_P},
leads to the initial value ${\rm P}_j(t) \simeq -\kappa\delta_{k,1} e^{\tau {\rm M}_{jk}}$ (summation over repeated indices is implied).
Now using the relationship between $\omega_{\rm a}$ and $p_\theta$ we finally get
\begin{equation}
    \omega_{\rm a}(t) = -\frac{p_{\theta}(t)}{2\mu s^3\gamma_r} \simeq \frac{\kappa}{2\mu s^3\gamma_r}e^{\tau {\rm M}_{31}} \, .
    \label{eq:omega_OC}
\end{equation}
Integrating Eqs.~(\ref{eq:app_motion2D_r},\ref{eq:app_motion2D_theta}) with Eq.~\eqref{eq:omega_OC}, we thus obtain an approximation of MDSP based on the local information about the environment stored in the coefficients of the matrix $\textbf{M}$.

Expanding the matrix exponential up to leading order terms in $\tau$, the policy~\eqref{eq:omega_OC} reduces to
\begin{equation}
    \omega_{\rm a} = -\frac{\kappa v_0 \tau}{2\mu s^3\gamma_r}\sin\theta + O(\tau^2) \, .
    \label{eq:omega_OC_approx}
\end{equation}
For small values of $\tau$, the policy amounts to assuming a uniform environment such that the swimmer points straight towards the finish line. 
On the other hand, the higher order contributions to~\eqref{eq:omega_OC_approx} depend on the flow structure and thus allow for smart navigation.

A local approximation of ZSP can be obtained similarly to the above derivation for MDSP.
As described in Table~\ref{tab:policies}, since in this case the swimmer is assumed point-like the state variable is the particle position $\bm r$ while the control is the steering direction $\bu$.
Applying OCT to this problem with the cost $\mathcal{C}_{\rm ZSP} = -\kappa(x(\ta) - x(t)) + \sigma(\ta - t)$, we obtain
\begin{equation}
        \dot{\bm{p}} = - \textbf{F}\cdot\bm{p}, \qquad
        \bu = -\frac{\bm{p}}{|\bm{p}|},
    \label{eq:ZSP_OC}
\end{equation}
with the boundary condition $\bm p(\ta) = -\kappa\hat{\bm x}$.
Following the same procedure that led from Eqs.~\eqref{eq:app_motion2D} to Eq.~\eqref{eq:omega_OC}, we assume the matrix $\textbf{F}$ to be constant over the time interval $[t;t + \tau]$, such that after solving for $\textbf{P}$ we get
\begin{equation}
    \bu(t) = \frac{e^{\tau \textbf{F}}\cdot\hat{\bm x}}{|e^{\tau \textbf{F}}\cdot\hat{\bm x}|} \, ,
    \label{eq:ZSP_overdamped}
\end{equation}
where $\textbf{F}$ is evaluated at time $t$.
As previously, the leading order contribution to Eq.~\eqref{eq:ZSP_overdamped} leads to $\bm u(t) = \hat{\bm x} + O(\tau)$, {\it i.e.} pointing straight at the finish line, while contributions from the flow show up at higher order.
We note that~\eqref{eq:ZSP_overdamped} was derived via a different method in Ref.~\cite{surfing}.

Eq.~\eqref{eq:ZSP_overdamped} describes instantaneous reorientations of the swimmer directions.
Hence, to compare the performances of ZSP and MDSP in Fig.~\ref{fig:3} we implemented an underdamped version of~\eqref{eq:ZSP_overdamped} obtained by simulating Eqs.~(\ref{eq:app_motion2D_r},\ref{eq:app_motion2D_theta}) with the control
\begin{equation}
    \omega_{{\rm a},{\rm ZSP}} = -\frac{v_0 \kappa \tau}{2\mu s^3\gamma_r}\sin(\theta - \theta_{\rm ZSP}) \, ,
    \label{eq:ZSP_underdamped}
\end{equation}
where $\theta_{\rm ZSP}$ denotes the orientation set by~\eqref{eq:ZSP_overdamped}.\\

\noindent \emph{Appendix on the numerical methods for the navigation in a Gaussian random flow.---}
The Gaussian random flow described in the main text was obtained via the power spectrum generation method~\cite{Goon}. 
We first build a $N \times N$ matrix of uncorrelated zero mean and unit variance Gaussian white noise, and then evaluate its Fourier transform.
%in Mathematica~\cite{mathematica}.
We multiply the outcome with the square root of the desired power spectrum of the stream function $\psi(\bm{r})$ ---the Fourier transform of its correlation function--- and Fourier transform the result back. 
Finally, the flow field is obtained using finite difference via ${\bm f}({\bm r}) = 2^{-1/2}\bm{\nabla}\times[\hat{\bm z} \psi(\bm{r})]$ where $\hat{\bm z}$ is the out-of-plane unit vector. The flow generated this way is by construction periodic across the domain boundaries. We used $N = 2500$ and a physical system size $L = 25 \ell$, leading to a spatial resolution of $\rmd x = 10^{-2}\ell$.

For simulations in random flow, the the equations of motion of the swimmer~\eqref{eq:steering} were numerically integrated together with the equation for the control ({\it i.e.} Eq.~\eqref{eq:omega_OC} for MDSP and Eq.~\eqref{eq:ZSP_underdamped} for ZSP) with a $4^{\rm th}$ order Runge-Kutta method and a time step ${\rm d}t=10^{-4}$.
The matrix exponentials in~\eqref{eq:omega_OC} and~\eqref{eq:ZSP_underdamped} were computed with the 
\texttt{gsl\_linalg\_exponential\_ss} routine included from the GSL library~\cite{gough2009}.

The results presented in the main text have been obtained from simulations of $N_{\rm ic}=10^3$ trajectories with initial positions ${\bm r}_0=y_0\hat{\bm y}$ and $y_0$ uniformly distributed in $\in[0, L]$. 
In all cases, the initial heading direction of the swimmer is set to $\theta_0=0$. 
The probability $\pi_{\rm fail}$ that a swimmer does not reach the finish line located at $x=4L$ is defined as the fraction of trajectories not crossing it within a time $t_{\rm max} = 5\times4L/v_0$. 
%Namely, $\pi_{\rm fail} \equiv N(\ta\ge t_{\rm max})/N_{\rm ic}$.
The mean and quartiles of both arrival time ($\ta$) and dissipated energy ($\cal C$) were computed considering only the trajectories that successfully reach the finish line, while the data points shown in Figs.~\ref{fig:3}(c,d) all satisfy $\pi_{\rm fail}\leq0.05$.

\bibliography{opt_shape}

\end{document}